\documentstyle[12pt]{article}
\begin{document}
\thispagestyle{empty}
\title{Interacting Spinor and Scalar Fields in Bianchi type-I Universe 
filled with Perfect Fluid: Exact Self-consistent Solutions}
\author{B. Saha\\ 
Laboratory of Theoretical Physics\\ 
Joint Institute for Nuclear Research, Dubna\\ 
141980 Dubna, Moscow region, Russia\\ 
e-mail:  saha@thsun1.jinr.dubna.su\\ 
G. N. Shikin\\ 
Department of Theoretical Physics\\ 
Russian Peoples' Friendship University\\ 
6, Miklukho-Maklay str., 117198 Moscow, Russia}
\date{}
\maketitle 
\thispagestyle{empty}

\noindent
Self-consistent solutions to interacting spinor and scalar field  equations 
in General Relativity are studied for the case of Bianchi type-I 
space-time filled with perfect fluid. The initial and the asymptotic
behavior of the field functions and the metric one has been thoroughly  
studied. 


\newpage
\section{Introduction}

\noindent
The quantum field theory in curved space-time has been a matter of
great interest in recent years because of its applications to 
cosmology and astrophysics. The evidence of existence of strong
gravitational fields in our Universe led to the study of the quantum 
effects of material fields in external classical gravitational field.
After the appearance of Parker's paper on scalar fields \cite{Par1} and 
spin-$\frac{1}{2}$ fields \cite{Par2}, several authors have studied this 
subject. Although the Universe seems homogenous and isotropic at present,
there is no observational data that guaranties the isotropy in the era prior
to the recombination. In fact, there are theoretical arguments that sustain 
the existence of an anisotropic phase that approaches an isotropic one 
\cite{Mis}. Interest in studying Klein-Gordon and Dirac equations in
anisotropic models has increased since Hu and Parker \cite{Hu0} have shown
that the creation of scalar particles in anisotropic backgrounds can
dissipate the anisotropy as the Universe expands. 

\noindent
A Bianchi type-I (B-I) Universe, being the straightforward generalization 
of the flat Robertson-Walker (RW) Universe, is one of the simplest models  
of an anisotropic Universe that describes a homogenous and spatially flat
Universe. Unlike the RW Universe which has the same 
scale factor for each of the three spatial directions, a B-I Universe
has a different scale factor in each direction, thereby introducing an
anisotropy to the system. It moreover has the agreeable property that
near the singularity it behaves like a Kasner Universe even in the 
presence of matter and consequently falls within the general analysis
of the singularity given by Belinskii et al \cite{Bel}. 
And in a Universe filled with matter for $p\,=\,\gamma\,\varepsilon, \quad 
\gamma < 1$, it has been shown that any initial anisotropy in a B-I
Universe quickly dies away and a B-I Universe eventually evolve
into a RW Universe \cite{Jac}. Since the present-day Universe is 
surprisingly isotropic, this feature of the B-I Universe makes it a prime 
candidate for studying the possible effects of an anisotropy in the early 
Universe on present-day observations. In light of the importance of
mentioned above, several authors have studied linear spinor field equations       
\cite{Chim}, \cite{Cas} and the behavior of 
gravitational waves (GW's) \cite{Hu}, \cite{Mied}, \cite{Cho}    
in B-I Universe. Nonlinear spinor field (NLSF) in external cosmological
gravitation field was first studied by G. N. Shikin in 1991  
\cite{Shik}. This study was extended by us for more general case where 
we consider nonlinear term as an arbitrary function of all possible 
invariants generated from spinor bilinear forms. In that paper we also
studied the possibility of elimination of initial singularity specially for
Kasner Universe \cite{Ryb1}. 
In a recent paper \cite{Ryb} we studied 
the behavior of self-consistent NLSF in B-I Universe that was 
followed by the papers \cite{Alv}, \cite{AlvIz} where we studied
the self-consistent system of interacting spinor and scalar fields.
The purpose of this paper is to extend our study for different kinds
of interacting term in presence of perfect fluid. In the section 2 we 
derive fundamental equations corresponding to the Lagrangian for the 
self-consistent system of spinor, scalar and gravitational fields 
in presence of perfect fluid and seek their general solutions. 
In section 3 we give a detail analysis of
the solutions obtained for different kinds of interacting term. 
In section 4 we sum up the results obtained.
\vskip 5mm
\section{Fundamental equations and general solutions}
\setcounter{equation}{0}
\noindent
The Lagrangian for the self-consistent system of  spinor  and gravitation 
fields in presence of perfect fluid is 
\begin{equation}
L\,=\,L_g + L_{sp} + L_{sc} + L_m + L_{int},
\end{equation}
where $L_g,\, L_{sp},\,L_{sc}$, corresponding to gravitational, 
free spinor and free scalar fields read
\begin{eqnarray} 
L_g&=& R/2\kappa, \nonumber \\
L_{sp}&=&(i/2) 
\bigl[ \bar \psi \gamma^{\mu} \nabla_{\mu} \psi- \nabla_{\mu} \bar 
\psi \gamma^{\mu} \psi \bigr] - m\bar \psi \psi, \nonumber \\
L_{sc}&=& (1/2) \varphi_{,\mu} \varphi^{,\mu}, \nonumber
\end{eqnarray} 
with $R$ being the scalar curvature and $\kappa$ being  the  
Einstein's gravitational constant and $L_m$ is the Lagrangian of 
perfect fluid. As interaction Lagrangian we consider the following cases
\cite{Alv}, \, \cite{AlvIz}, \, \cite{Schw}~:
\begin{eqnarray}
1.\,\, L_{int}&=& (\lambda/2)\,\varphi_{,\alpha}\varphi^{,\alpha} F,
\nonumber \\
2.\,\, L_{int}&=& \lambda \bar \psi \gamma^\mu \psi \varphi_{,\mu},  
\nonumber \\
3.\,\, L_{int}&=& i \lambda \bar \psi \gamma^\mu \gamma^5 \psi \varphi_{,\mu},
\nonumber
\end{eqnarray}
where $\lambda$ is the coupling constant and $F$ can be presented
as some arbitrary functions of invariants generated from the real bilinear 
forms of spinor field having the form: 
$$S\,=\, \bar \psi \psi, \quad 
P\,=\,i \bar \psi \gamma^5 \psi, \quad
v^\mu\,=\,(\bar \psi \gamma^\mu \psi), \quad
A^\mu\,=\,(\bar \psi \gamma^5 \gamma^\mu \psi), \quad
T^{\mu\nu}\,=\,(\bar \psi \sigma^{\mu\nu} \psi),$$
where $\sigma^{\mu\nu}\,=\,(i/2)[\gamma^\mu\gamma^\nu\,-\,
\gamma^\nu\gamma^\mu]$. Invariants, corresponding to the bilnear forms, look 
$$ I = S^2, \quad J = P^2, \quad 
I_A = A_\mu\,A^\mu\,=\,(\bar \psi \gamma^5 \gamma^\mu \psi)\,g_{\mu\nu}
(\bar \psi \gamma^5 \gamma^\nu \psi),$$ 
$$ I_v = v_\mu\,v^\mu\,=\,(\bar \psi \gamma^\mu \psi)\,g_{\mu\nu}
(\bar \psi \gamma^\nu \psi),\quad
I_T = T_{\mu\nu}\,T^{\mu\nu}\,=\,(\bar \psi \sigma^{\mu\nu} \psi)\,
g_{\mu\alpha}g_{\nu\beta}(\bar \psi \sigma^{\alpha\beta} \psi).$$ 
According to the Pauli-Fierz theorem \cite{Ber} among the five invariants
only $I$ and $J$ are independent as all other can be expressed by them:
$$I_v = - I_A = I + J, \qquad I_T = I - J.$$ 
Therefore we choose $F = F(I, J)$.
 
\noindent
We choose B-I space-time metric in the form 
\begin{equation}
ds^2\,=\,dt^2 - \gamma_{ij}(t)\,dx^i\,dx^j.
\end{equation}
As it admits no rotational matter, the spatial metric $\gamma_{ij}(t)$
can be put into diagonal form. Now we can rewrite the B-I space-time metric 
in the form \cite{Zel}:
\begin{equation} ds^2 = dt^2 - a^{2}(t)\,dx^{2} - b^{2}(t)\,dy^{2} -
c^{2}(t)\,dz^2,
\end{equation}
where the velocity of light c is taken to be unity. 
Let us now write the Einstein equations for $a(t), b(t)$ and $c(t)$ 
corresponding to the metric (2.3) and Lagrangian (2.1)~\cite{Zel}: 
\begin{eqnarray}
\frac{\ddot a}{a} +\frac{\dot a}{a} \biggl(\frac{\dot b}{b}+\frac{\dot 
c}{c}\biggr)= -\kappa \biggl(T_{1}^{1}- \frac{1}{2}T\biggr), \\
\frac{\ddot b}{b} +\frac{\dot b}{b} \biggl(\frac{\dot a}{a}+\frac{\dot 
c}{c}\biggr)= -\kappa \biggl(T_{2}^{2}- \frac{1}{2}T\biggr),  \\
\frac{\ddot c}{c} +\frac{\dot c}{c} \biggl(\frac{\dot a}{a}+\frac{\dot 
b}{b}\biggr)= -\kappa \biggl(T_{3}^{3}- \frac{1}{2}T\biggr),   \\
\frac{\ddot a}{a} +\frac{\ddot b}{b} +\frac{\ddot 
c}{c}= -\kappa \biggl(T_{0}^{0}- \frac{1}{2}T\biggr),
\end{eqnarray}
where points denote differentiation with respect to t, and $T_{\mu}^{\nu}$
is the energy-momentum tensor of material fields and perfect fluid.  

\noindent
The scalar and the spinor field equations and the energy-momentum tensor of 
material fields and perfect fluid corresponding to (2.1) are
\begin{equation}
\partial_\alpha\, \bigl[\sqrt{-g}\,(g^{\alpha \beta} \varphi_{,\beta} +
\partial L_{int} /\partial \varphi_{,\alpha}) \bigr]\,=\,0,
\end{equation}
\begin{eqnarray}
i\gamma^\mu \nabla_\mu \psi -m\psi + \partial L_{int} / \partial \bar \psi
&=&0, \nonumber \\
i \nabla_\mu \bar \psi \gamma^\mu + m \bar\psi - 
\partial L_{int} / \partial \psi &=&0. 
\end{eqnarray}
\begin{eqnarray}
T_{\mu}^{\rho}&=&\frac{i}{4} g^{\rho\nu} \biggl(\bar \psi \gamma_\mu 
\nabla_\nu \psi + \bar \psi \gamma_\nu \nabla_\mu \psi - \nabla_\mu \bar 
\psi \gamma_\nu \psi - \nabla_\nu \bar \psi \gamma_\mu \psi \biggr) 
\nonumber \\
&+&\varphi_{,\mu} \varphi^{,\rho} + 2 \frac{\delta L_{int}}
{\delta g^{\mu\rho}} g^{\rho \nu}
-\delta_{\mu}^{\rho} (L_{sp} + L_{sc}
+ L_{int}) + T_{\mu\,(m)}^{\rho}.
\end{eqnarray}
Here $T_{\mu\,(m)}^{\rho}$ is the 
energy-momentum tensor of perfect fluid. For a Universe filled with
perfect fluid, in the concomitant system of reference 
$(u^0=1, \, u^i=0, i=1,2,3)$ we have
\begin{equation}
T_{\mu (m)}^{\nu}\,=\, (p + \varepsilon) u_\mu u^\nu - 
\delta_{\mu}^{\nu} p \,=\,(\varepsilon,\,- p,\,- p,\,- p),
\end{equation} 
where energy $\varepsilon$ is related to the pressure $p$ by the
equation of state $p\,=\,\gamma\,\varepsilon$, the general solution
has been derived by Jacobs \cite{Jac}. $\gamma$ varies between the
interval $0\,\le\, \gamma\,\le\,1$, whereas $\gamma\,=\,0$ describes
the dust Universe, $\gamma\,=\,\frac{1}{3}$ presents radiation Universe,
$\frac{1}{3}\,<\,\gamma\,<\,1$ ascribes hard Universe and $\gamma\,=\,1$
corresponds to the stiff matter. As one sees changes in the solutions 
performed by perfect fluid carried out through Einstein equations, namely 
through $\tau  = a(t) b(t) c(t)$. So, let us first see how the 
quantities $\varepsilon$ and $p$ connected with $\tau$.
In doing this we use the well-known equality $T_{\mu;\nu}^{\nu}\,=\,0$,
that leads to
\begin{equation}
\frac{d}{dt}(\tau \varepsilon) + {\dot \tau} p\,=\,0,
\end{equation}
with the solution
\begin{equation}
\mbox{ln} \tau\,=\,-\int\,\frac{d \varepsilon}{(\varepsilon + p)}.
\end{equation}
Recalling the equation of state 
$p\,=\,\xi \varepsilon,\,\, 0 \le \xi \le 1$ finally we get
\begin{equation}
T_{0 (m)}^{0}\,=\,\varepsilon\,=\,\frac{\varepsilon_0}{\tau^{1+\xi}}, \,\,
T_{1 (m)}^{1}\,=\,T_{2 (m)}^{2}\,=\,T_{3 (m)}^{3}\,=\,- p\,=\,-
\frac{\varepsilon_0 \xi}{\tau^{1+\xi}},
\end{equation}
where $\varepsilon_0$ is the integration constant. 

\noindent
Note that we consider space-independent field only. Under this assumption
and with regard to spinor field equations, the components of the 
energy-momentum tensor read:
\begin{equation}
T_{0}^{0}=mS+ \frac{1}{2}\dot{\varphi}^2 + L_{int} + \varepsilon, \quad
T_{1}^{1}=T_{2}^{2}=T_{3}^{3}=\frac{1}{2}\bigl(\bar\psi 
\frac{\partial L_{int}}{\partial \bar \psi} +
\frac{\partial L_{int}}{\partial \psi} \psi\bigr) - L_{sc} - L_{int} -p.
\end{equation} 

\noindent
In (2.8) and (2.10) $\nabla_\mu$ denotes the covariant 
derivative of spinor, having the form \cite{Zelnor}:  
\begin{equation} \nabla_\mu 
\psi=\frac{\partial \psi}{\partial x^\mu} -\Gamma_\mu \psi, \end{equation} 
where $\Gamma_\mu(x)$ are spinor affine connection matrices.  
$\gamma^\mu(x)$ matrices are defined for the metric (2.3) as follows. 
Using the equalities  \cite{Brill}, \cite{Wein}
$$ g_{\mu \nu} (x)= e_{\mu}^{a}(x) e_{\nu}^{b}(x) \eta_{ab},
\qquad \gamma_\mu(x)\,=\,e_{\mu}^{a}(x)\bar\gamma^a,$$ 
where $\eta_{ab}= \mbox{diag}(1,-1,-1,-1)$,
$\bar \gamma_\alpha$ are the Dirac matrices of Minkowski space and
$e_{\mu}^{a}(x)$ are the set of tetradic 4-vectors, we obtain 
the Dirac matrices $\gamma^\mu(x)$ of curved space-time
$$ \gamma^0=\bar \gamma^0,\quad \gamma^1 =\bar \gamma^1 /a(t),\quad 
\gamma^2= \bar \gamma^2 /b(t),\quad \gamma^3 = \bar \gamma^3 /c(t), $$
$$ \gamma_0=\bar \gamma_0,\quad \gamma_1 =\bar \gamma_1 a(t),\quad 
\gamma_2= \bar \gamma_2 b(t),\quad \gamma_3 = \bar \gamma_3 c(t). $$
$\Gamma_\mu(x)$ matrices are defined by the equality $$\Gamma_\mu (x)= 
\frac{1}{4}g_{\rho\sigma}(x)\biggl(\partial_\mu e_{\delta}^{b}e_{b}^{\rho} 
- \Gamma_{\mu\delta}^{\rho}\biggr)\gamma^\sigma\gamma^\delta, $$ 
which gives
\begin{equation} \Gamma_0=0, \quad \Gamma_1=\frac{1}{2}\dot a(t) 
\bar \gamma^1 \bar \gamma^0, \quad \Gamma_2=\frac{1}{2}\dot b(t) \bar 
\gamma^2 \bar \gamma^0, \quad \Gamma_3=\frac{1}{2}\dot c(t) \bar \gamma^3 
\bar \gamma^0.\end{equation}
Flat space-time matrices we choose in the form, given in 
\cite{Bog}:
\begin{eqnarray}
\bar \gamma^0&=&\left(\begin{array}{cccc}1&0&0&0\\0&1&0&0\\
0&0&-1&0\\0&0&0&-1\end{array}\right), \quad
\bar \gamma^1\,=\,\left(\begin{array}{cccc}0&0&0&1\\0&0&1&0\\
0&-1&0&0\\-1&0&0&0\end{array}\right), \nonumber\\
\bar \gamma^2&=&\left(\begin{array}{cccc}0&0&0&-i\\0&0&i&0\\
0&i&0&0\\-i&0&0&0\end{array}\right), \quad
\bar \gamma^3\,=\,\left(\begin{array}{cccc}0&0&1&0\\0&0&0&-1\\
-1&0&0&0\\0&1&0&0\end{array}\right).  \nonumber
\end{eqnarray}
Defining $\gamma^5$ as follows
\begin{eqnarray}
\gamma^5&=&-\frac{i}{4} E_{\mu\nu\sigma\rho}\gamma^\mu\gamma^\nu
\gamma^\sigma\gamma^\rho, \quad E_{\mu\nu\sigma\rho}= \sqrt{-g}
\varepsilon_{\mu\nu\sigma\rho}, \quad \varepsilon_{0123}=1,\nonumber \\
\gamma^5&=&-i\sqrt{-g} \gamma^0 \gamma^1 \gamma^2 \gamma^3 
\,=\,-i\bar \gamma^0\bar \gamma^1\bar \gamma^2\bar \gamma^3 =
\bar \gamma^5, \nonumber
\end{eqnarray}
we obtain
\begin{eqnarray}
\bar \gamma^5&=&\left(\begin{array}{cccc}0&0&-1&0\\0&0&0&-1\\
-1&0&0&0\\0&-1&0&0\end{array}\right).\nonumber
\end{eqnarray}

\noindent
Let us now solve the Einstein equations. With respect to (2.15)
summation of Einstein equations (2.4), (2.5) and (2.6) leads to the equation 
\begin{equation}
\frac{\ddot 
\tau}{\tau}=-\kappa(T_{1}^{1}+T_{2}^{2}+T_{3}^{3}-\frac{3}{2}T)=
\frac{3 \kappa}{2} \,(T_{0}^{0} + T_{1}^{1}).
\end{equation} 
In case if the right hand side of (2.18) be the 
function of $\tau(t)\,=\,a(t)b(t)c(t)$, this equation takes the form 
\begin{equation}
\ddot \tau+\Phi(\tau)=0.            
\end{equation}
As is known this equation possesses exact solutions  for  
arbitrary  function $\Phi(\tau)$.  Giving the explicit 
form of $L_{int}$, from (2.18)  one  can find concrete 
function $\tau(t)=abc$. 
Once the value of $\tau$ is obtained, one can 
get expressions for components $V_\alpha(t), \quad \alpha =1,2,3,4.$
Let us express $a, b, c$ through $\tau$. For this we notice that
subtraction of Einstein equations  (2.4) - (2.5)  leads  to  
the equation 
\begin{equation}
\frac{\ddot a}{a}-\frac{\ddot b}{b}+\frac{\dot a \dot c}{ac}- 
\frac{\dot b \dot c}{bc}= \frac{d}{dt}\biggl(\frac{\dot a}{a}- 
\frac{\dot b}{b}\biggr)+\biggl(\frac{\dot a}{a}- \frac{\dot b}{b} \biggr) 
\biggl (\frac{\dot a}{a}+\frac{\dot b}{b}+ \frac{\dot c}{c}\biggr)= 0. 
\end{equation} 
Equation (2.20) possesses the solution
\begin{equation}
\frac{a}{b}= D_1 \mbox{exp} \biggl(X_1 \int \frac{dt}{\tau}\biggr), \quad 
D_1=\mbox{const.}, \quad X_1= \mbox{const.} \end{equation}
Subtracting equations (2.4) - (2.6) and (2.5) - (2.6) one finds   
the equations similar to (2.20), having solutions \begin{equation} 
\frac{a}{c}= D_2 \mbox{exp} \biggl(X_2 \int \frac{dt}{\tau}\biggr), \quad 
\frac{b}{c}= D_3 \mbox{exp} \biggl(X_3 \int \frac{dt}{\tau}\biggr),  
\end{equation}
where $D_2, D_3, X_2, X_3 $ are integration constants. There is a 
functional dependence between the constants 
$D_1,\, D_2,\, D_3,\, X_1,\, X_2,\, X_3 $:  
$$ D_2=D_1\, D_3, \qquad X_2= X_1\,+\,X_3.$$
Using the equations (2.21) and (2.22), we rewrite $a(t), b(t), c(t)$ in 
the explicit form:  
\begin{eqnarray} a(t) &=& 
(D_{1}^{2}D_{3})^{\frac{1}{3}}\tau^{\frac{1}{3}}\mbox{exp}\biggl[\frac{2X_1 
+X_3}{3} \int\,\frac{dt}{\tau (t)} \biggr], \nonumber \\
b(t) &=& 
(D_{1}^{-1}D_{3})^{\frac{1}{3}}\tau^{\frac{1}{3}}\mbox{exp}\biggl[-\frac{X_1 
-X_3}{3} \int\,\frac{dt}{\tau (t)} \biggr], \nonumber \\
c(t) &=& 
(D_{1}D_{3}^{2})^{-\frac{1}{3}}\tau^{\frac{1}{3}}\mbox{exp}\biggl[-\frac{X_1 
+2X_3}{3} \int\,\frac{dt}{\tau (t)} \biggr]. 
\end{eqnarray}
Thus the previous system of Einstein equations is completely integrated.  
In this process of integration only first three of the complete system of 
Einstein equations have been used. General solutions to these three second 
order equations have been obtained. The  solutions  contain  six arbitrary 
constants: $D_1, D_3, X_1, X_3 $  and two others, 
that  were obtained while solving  equation  (2.19).  
Equation  (2.7)  is  the consequence of first three of Einstein equations.  
To  verify  the correctness of obtained solutions, it is necessary to put 
$a, b, c$ into (2.7). It should lead either to identity or to  some  
additional constraint between the constants. Putting $a, b, c$  from  (2.23) 
into (2.7) one can get the following equality:  
\begin{equation}
\frac{\ddot \tau}{\tau} - \frac{2}{3}\,\frac{\dot{\tau}^2}{\tau^2} +
\frac{2}{9 \tau^2} {\cal X} \, = \, - \frac{\kappa}{2}\, (T_{0}^{0} -
3 T_{1}^{1}), \quad {\cal X} := X_{1}^{2} + X_1 X_3 + X_{3}^{2},
\end{equation}
that guaranties the correctness of the solutions obtained.
This together with (2.18) gives the equation for $\tau$ with the
solution in quadrature:
\begin{equation}
\int\,\frac{d\tau}{\sqrt{3\kappa \tau^2 T_{0}^{0} + {\cal X}/3}}\,=\,t.
\end{equation}  

\noindent
It should be emphasized that we are dealing with cosmological problem
and our main goal is to investigate the initial and the asymptotic behavior
of the field functions and the metric ones. As one sees, all these functions
are in some functional dependence with $\tau$. Therefore in our further
investigation we mainly look for $\tau$, though in some particular cases
we write down field and metric functions explicitly.

\vskip 5mm
\section{Analysis of the solutions obtained for some special choice
of interaction Lagrangian}
\setcounter{equation}{0}
\noindent
Let us now study the system for some special choice of $L_{int}$.  
We first study the solution to the system of field equations with 
minimal coupling when the direct interaction between the spinor and 
scalar fields remains absent.
The reason to get the solution  to the self-consistent system of 
equations for the fields with minimal coupling is the necessity 
of comparing this solution with that for the  system  of equations   for 
the interacting spinor, scalar and gravitational fields  that permits  to 
clarify the  role of interaction terms in the evolution of the 
cosmological model in question. 

\noindent
In this case from the scalar and spinor field equations one finds 
$\dot \varphi = C/\tau$ and $\bar \psi \psi = S = C_0/\tau$ with
$C$ and $C_0$ being the constants of integration. Therefore
the components of the energy-momentum tensor look:
\begin{eqnarray}
T_{0}^{0} = \frac{mC_0}{\tau}+ \frac{C^2}{2 \tau^2}, \quad 
T_{1}^{1}=T_{2}^{2}=T_{3}^{3}=- \frac{C^2}{2 \tau^2}.  
\end{eqnarray}
Note that as the energy density $T_{0}^{0}$ should be a quantity 
positively defined, the equation (3.1) leads to $C_0 >0$. The inequality 
$C_0 >0$ will also be preserved for the system with direct interaction 
between the fields as in this case the correspondence principle should be 
fulfilled: for $\lambda =0$ the field system with direct interaction turns 
into that with minimal coupling.

\noindent
The components of spinor field functions in this case read
\begin{equation} \psi_{1,2}(t) = 
(C_{1,2}/\sqrt{\tau})\,e^{-imt}, \quad \psi_{3,4}(t) = 
(C_{3,4}/\sqrt{\tau})\,e^{imt}. \end{equation} 
Taking into account (3.1) equation (2.25) writes 
\begin{equation}
\int\,\frac{d\tau}{\sqrt{3\kappa m C_0 \tau + 3 \kappa C^2/2 + 
{\cal X}/3}}\,=\,t.
\end{equation}
with the solution
$$ \tau|_{t \to 0} \approx \sqrt{3 \kappa C^2/2 + {\cal X}/3}\, t \to 0,$$
and 
$$ \tau|_{t \to \infty} \approx \sqrt{3 \kappa m C_0}\,t^2.$$
Thus one concludes that the solutions obtained are initially singular 
and the space-time is asymptotically isotropic.

\noindent
Let us now study the case with different kinds of interactions.

\noindent
{\bf Case 1.} For the case when $L_{int}= (\lambda/2) \varphi_{,\mu}
\varphi^{,\mu} F(I, J)$ one writes the scalar field equation as
\begin{equation}
\frac{\partial}{\partial t} \bigl(\tau \dot \varphi (1 + \lambda F)\bigr) = 0,
\end{equation} 
with the solution 
\begin{equation}
\dot{\varphi}\,=\,C /\tau (1 + \lambda F). 
\end{equation}
In this case the first equation of the system (2.9) now reads 
\begin{equation} i\bar \gamma^0 
\biggl(\frac{\partial}{\partial t} +\frac{\dot \tau}{2 \tau} \biggr) \psi 
-m \psi +{\cal D} \psi + i {\cal G} \gamma^5\psi=0,   
\end{equation}
where ${\cal D} := \, \varphi_{,\alpha} \varphi^{,\alpha} S\, F_I$ and
${\cal G}:=\, \varphi_{,\alpha} \varphi^{,\alpha} P\, F_J.$ For the 
components $\psi_\rho= V_\rho(t)$, where $\rho=1,2,3,4,$ from (3.6) 
one deduces the following system of equations:  
\begin{eqnarray} 
{\dot V}_1 +\frac{\dot \tau}{2 \tau} V_1 
+i(m- {\cal D}) V_1 - {\cal G}V_3 &=& 0,  \nonumber\\
{\dot V}_2 +\frac{\dot \tau}{2 \tau} V_2 
+i(m- {\cal D}) V_2 - {\cal G}V_4 &=& 0,  \nonumber\\
{\dot V}_3 +\frac{\dot \tau}{2 \tau} V_3 
-i(m- {\cal D}) V_3 + {\cal G}V_1 &=& 0,  \nonumber \\
{\dot V}_4 +\frac{\dot \tau}{2 \tau} V_4 
-i(m- {\cal D}) V_4 + {\cal G}V_2 &=& 0.
\end{eqnarray} 

\noindent
Let us now define the equations for
\begin{eqnarray}
P\,=\,i(V_1 V_{3}^{*}-V_{1}^{*}V_3 +V_2V_{4}^{*}-V_{2}^{*}V_4), \nonumber \\
R\,=\,(V_1 V_{3}^{*}+V_{1}^{*}V_3 +V_2V_{4}^{*}+V_{2}^{*}V_4), \nonumber \\
S\,=\,(V_{1}^{*} V_{1}+V_{2}^{*}V_2 -V_{3}^{*}V_{3}-V_{4}^{*}V_4). 
\end{eqnarray}
After a little manipulation one finds
\begin{eqnarray}
\frac{d S_0}{d t} -2 {\cal G}\, R_0\,=\,0, \nonumber\\
\frac{d R_0}{d t}+2 (m- {\cal D})\, P_0 + 2 {\cal G} S_0\,=\,0, \nonumber\\ 
\frac{d P_0}{d t}-2 (m- {\cal D})\, R_0\,=\,0,
\end{eqnarray}
where $S_0 = \tau S, \quad P_0 = \tau P, \quad R_0 = \tau R$.
From this system one can easily find
\begin{eqnarray}
S_0 {\dot S}_0 + R_0 {\dot R}_0 +P_0 {\dot P}_0\,=\,0, \nonumber
\end{eqnarray}
that gives
\begin{equation}
S^2 + R^2 + P^2 \,=\, A^2/ \tau^2, \qquad A^2 = \mbox{const.}
\end{equation}

\noindent
Let us go back to the system of equations (3.7). It can be written as
follows if one defines $W_\alpha\,=\,\sqrt{\tau}\,V_\alpha$:
\begin{eqnarray} 
{\dot W}_1 +i\Phi W_1 - {\cal G}W_3 &=& 0, \quad
{\dot W}_2 +i\Phi W_2 - {\cal G}W_4 = 0, \nonumber\\
{\dot W}_3 -i\Phi W_3 + {\cal G}W_1 &=& 0, \quad
{\dot W}_4 -i\Phi W_4 + {\cal G}W_2 = 0,
\end{eqnarray} 
where $\Phi\,=\,m- {\cal D}$. Defining  $U(\sigma) = W (t)$, where 
$\sigma = \int\,{\cal G} dt$, we rewrite the foregoing system as:
\begin{eqnarray} 
U_{1}^{\prime} + i (\Phi/{\cal G}) U_{1} - U_{3} &=& 0, \qquad
U_{2}^{\prime} + i (\Phi/{\cal G}) U_{2} - U_{4} = 0, \nonumber\\
U_{3}^{\prime} - i (\Phi/{\cal G}) U_{3} + U_{1} &=& 0, \qquad
U_{4}^{\prime} - i (\Phi/{\cal G}) U_{4} + U_{2} = 0,
\end{eqnarray} 
where  prime ($^\prime$) denotes differentiation with respect to $\sigma$.
One can now define $V_\alpha$ giving the explicit value of $L_{int}$.

\noindent
{\bf I.} Let us consider the case when $F\,=\,I^{n}\,=\,S^{2n}$.
It is clear that in this case ${\cal G}\,=\,0$.
From (3.9) we find  
\begin{equation}
S = C_0/\tau, \quad C_0= \mbox{const.}
\end{equation}
As in the considered case $F$ depends only on $S$, from (3.13) it follows 
that ${\cal D}$ is a functions of $\tau=abc$. Taking this fact into account, 
integration of the system of equations (3.11) leads to the expressions 
\begin{eqnarray} 
V_{r}(t) = (C_r/\sqrt{\tau})\,e^{-i\Omega}, \quad r=1,2, \quad 
V_{l}(t) = (C_l/\sqrt{\tau})\,e^{i\Omega}, \quad l=3,4,  
\end{eqnarray} where $ C_r$ and 
$C_l$ are integration constants and $\Omega =\int\,\Phi (t) dt.$
Putting this solution into (3.8) one gets
\begin{equation}
S\,=\,(C_{1}^{2} + C_{2}^{2} - C_{3}^{2} - C_{4}^{2})/\tau.
\end{equation}
Comparing it with (3.13) we find $C_0 = 
C_{1}^{2} + C_{2}^{2} - C_{3}^{2} - C_{4}^{2}.$
The equation (2.25) in this case reads
\begin{equation} \int \frac{d \tau}{\sqrt{3 \kappa \bigl(mC_0 \tau +
\frac{C^2}{2(1 + \lambda\, C_{0}^{2n}/ \tau^{2n})} + 
\varepsilon_0 \tau^{1 -\xi}\bigr)+ {\cal X}/3}}= t. 
\end{equation} 
As one sees
$$ 
\tau(t)|_{t \to \infty} \approx \frac{3}{4}\kappa m C_0 t^2 \to \infty,$$ 
$$ \tau(t)|_{t \to 0} \approx \sqrt{{\cal X}/3}\,t \to 0.$$
Thus in the case considered, the asymptotical isotropization of the expansion 
process of initially anisotropic Bianchi type-I space-time takes place 
without the influence of scalar field. For a detail analysis of this case
see \cite{Alv}.
\vskip 3mm
\noindent
{\bf II.} We study the system when $F\,=\,J^{n}\,=\,P^{2n}$,
which means in the case considered ${\cal D}\,=\,0$. Let us 
note that, the interaction between the fields inevitably leads to the  
appearnce of nonlinear terms in the field equations. As is known,        
in the unified nonlinear spinor theory of Heisenberg the 
massive term remains absent, as according to Heisenberg, the particle 
mass should be obtained as a result of quantization of spinor prematter 
\cite{Hei}. In nonlinear generalization of classical field equations, 
the massive term does not possess the significance that it possesses 
in linear one, as by no means it defines total energy (or mass) of 
nonlinear field system. Thus without losing the generality we can 
consider massless spinor field putting $m\,=\,0$ that leads to 
$\Phi\,=\,0.$ Then from (3.9) one gets
\begin{equation}
P(t)\,=\,D_0/\tau, \,\, D_0=\,\mbox{const.}
\end{equation}
The system of equations (3.12) in this case reads
\begin{eqnarray} 
U_{1}^{\prime} -  U_{3} &=& 0, \qquad
U_{2}^{\prime} -  U_{4} = 0, \nonumber\\
U_{3}^{\prime} +  U_{1} &=& 0, \qquad
U_{4}^{\prime} +  U_{2} = 0.
\end{eqnarray} 
Differentiating the first equation of system (3.18) and taking into 
account the third one we get 
\begin{equation}
U_{1}^{\prime \prime} +U_{1} =\,0,
\end{equation}
which leads to the solution
\begin{equation}
U_1 = D_1 e^{i \sigma} + iD_3 e^{-i \sigma},\quad
U_3 =  i D_1 e^{i \sigma} + D_3 e^{-i \sigma}.
\end{equation}
Analogically for $U_2$ and $U_4$ one gets
\begin{equation}
U_2 = D_2 e^{i \sigma} + iD_4 e^{-i \sigma},\quad
U_4 = i D_2 e^{i \sigma} + D_4 e^{-i \sigma}, 
\end{equation}
where $D_i$ are the constants of integration.
Finally, we can write
\begin{eqnarray}
V_1&=&(1/\sqrt{\tau}) (D_1 e^{i \sigma} + iD_3 
e^{-i\sigma}), \quad
V_2 = (1/\sqrt{\tau}) (D_2 e^{i \sigma} + iD_4
e^{-i\sigma}),
\nonumber \\
V_3&=&(1/\sqrt{\tau}) (iD_1 e^{i \sigma} + D_3
e^{-i \sigma}), \quad
V_4 = (1/\sqrt{\tau}) (iD_2 e^{i \sigma} + D_4
e^{-i\sigma}).
\end{eqnarray} 
Putting (3.22) into (3.8) one finds
\begin{equation}
P=2\,(D_{1}^{2} + D_{2}^{2} - D_{3}^{2} -D_{4}^{2})/\tau.
\end{equation}
Comparison of (3.17) with (3.23) gives
$D_0=2\,(D_{1}^{2} + D_{2}^{2} - D_{3}^{2} -D_{4}^{2}).$
Let us now estimate $\tau$ using the equation
\begin{equation} \int \frac{d \tau}{\sqrt{3\kappa\bigl( 
\frac{C^2}{2(1 + \lambda\, C_{0}^{2n}/ \tau^{2n})} + 
\varepsilon_0 \tau^{1 -\xi}\bigr)+ {\cal X}/3}}= t. 
\end{equation} 
In this case we obtain
$$ \tau|_{t \to \infty} \approx \bigl([\sqrt{\varepsilon_0}(\xi +1)/2]\,
t\bigr)^{2/(\xi + 1)}, $$
$$ \tau|_{t \to 0} \approx \sqrt{{\cal X}/3}\, t,$$
i.e. The solutions obtained are initially singular and the space-time is
asymptotically isotropic if $\xi < 1$ and anisotropic if $\xi = 1.$
\vskip 3mm
\noindent
{\bf III.} In this case we study $F\,=\,F(I,\,J)$. Choosing 
\begin{equation}
F\,=\,F(K_{\pm}), \quad K_{+} = I + J = I_v = -I_A, \quad
K_{-} = I - J = I_T,
\end{equation}
in case of massless spinor field we find
$$
{\cal D}\,=\,\varphi_{,\mu} \varphi^{,\mu} S F_{K_{\pm}}, \quad
{\cal G}\,=\,\pm\varphi_{,\mu} \varphi^{,\mu} S F_{K_{\pm}}, \quad
F_{K_{\pm}} = dF/dK_{\pm}.$$
Putting them into (3.9) we find
\begin{equation}
S_{0}^{2} \pm P_{0}^{2} = D_{\pm}.
\end{equation}
Choosing $F \,=\, K_{\pm}^{n}$ from (2.25) one comes to the 
conclusion similar to that of previous case ({\bf II}).

\noindent
{\bf Case 2.} In this case the scalar and spinor field equations read
\begin{equation}
\frac{\partial}{\partial t} \bigl[\tau\,(\dot \varphi +
\lambda \bar \psi \gamma^0 \psi) \bigr]\,=\,0,
\end{equation}
\begin{eqnarray}
i\gamma^0 \bigl(\frac{\partial}{\partial t} + \frac{\dot \tau}{2 \tau}
\bigr) \psi -m\psi + \lambda \dot \varphi \gamma^0 \psi
&=&0, \nonumber \\
i \bigl(\frac{\partial}{\partial t} + \frac{\dot \tau}{2 \tau}
\bigr) \bar \psi \gamma^0 + m \bar \psi - \lambda \dot \varphi 
\bar \psi \gamma^0 &=&0. 
\end{eqnarray}
Using the spinor field equations one finds: 
$\bar \psi \gamma^0 \psi = C_1/\tau,$ and $S=\bar \psi \psi = C_0/\tau$
with $C_1$ and $C_0$ being the constant of
integration. Putting it in the scalar field equation one obtains
\begin{equation}
\dot \varphi = (C - \lambda C_1)/\tau, \quad C = \rm{const.}
\end{equation} 
In account with all these the spinor field equation can be written as
\begin{equation}
\gamma^0 \bigl(\frac{\partial}{\partial t} +\frac{\dot \tau}{2 \tau}\bigr)
\psi + im \psi - \frac{i \lambda (C -\lambda C_1)}{\tau} \gamma^0 \psi =0,
\end{equation}
with the solution
\begin{eqnarray}
\psi_{1,2}(t) &=& \frac{D_{1,2}}{\sqrt{\tau}} \mbox{exp}\,
\bigr[-i\{mt -\lambda (C -\lambda C_1)\int\,\tau^{-1} dt\}\bigr],
\nonumber \\
\psi_{3,4}(t) &=& \frac{D_{3,4}}{\sqrt{\tau}} \mbox{exp}\,
\bigl[i\{mt +\lambda (C -\lambda C_1)\int\,\tau^{-1} dt\}\bigr].
\end{eqnarray}
The components of energy-momentum tensor in this case read
$$ T_{0}^{0} = \frac{m C_0}{\tau} + \frac{{\cal C}}{\tau^2} + 
\frac{\varepsilon_0}{\tau^{1 + \xi}}, \quad 
T_{1}^{1} = T_{2}^{2} = T_{3}^{3} = -\frac{(C - \lambda C_1)^2}{2 \tau^2}
- \frac{\varepsilon_0 \xi}{\tau^{1 +\xi}},$$
where ${\cal C} := (C^2 - \lambda^2 C_{1}^{2})/2$ and
$0<\xi<1.$ Putting this into (2.25) one gets
\begin{equation}
\int\,\frac{d\tau}{\sqrt{3\kappa (m C_0 \tau + \varepsilon_0 \tau^{1-\xi} 
+{\cal C})+ {\cal X}/3}} = t.
\end{equation}
As one sees
$$ \tau|_{t \to 0} \approx \sqrt{{\cal X}/3 - 3\kappa {\cal C}}\, t \to 0,$$
and
$$ \tau|_{t \to \infty} \approx \sqrt{3 \kappa m C_0}\, t^2,$$
which means the solution obtained is initially singular and the 
isotropization process of the initially anisotropic Universe takes place
as $t \to \infty.$

\noindent
{\bf Case 3.} In this case the scalar and spinor field equations read
\begin{equation}
\frac{\partial}{\partial t} \bigl[\tau\,(\dot \varphi +
i\lambda \bar \psi \gamma^0 \gamma^5 \psi) \bigr]\,=\,0,
\end{equation}
\begin{eqnarray}
i\gamma^0 \bigl(\frac{\partial}{\partial t} + \frac{\dot \tau}{2 \tau}
\bigr) \psi -m\psi + i\lambda \dot \varphi \gamma^0 \gamma^5 \psi
&=&0, \nonumber \\
i \bigl(\frac{\partial}{\partial t} + \frac{\dot \tau}{2 \tau}
\bigr) \bar \psi \gamma^0 + m \bar \psi - \lambda \dot \varphi 
\bar \psi \gamma^0 \gamma^5 &=&0. 
\end{eqnarray}
We consider the massless spinor field. In this case from the spinor 
field equations one finds: 
$i \bar \psi \gamma^0 \gamma^5 \psi = C_2/\tau,$ 
with $C_2$ being the constant of
integration. Putting it in the scalar field equation one obtains
\begin{equation}
\dot \varphi = (C - \lambda C_2)/\tau, \quad C = \rm{const.}
\end{equation} 
In account with all these the spinor field equation can be written as
\begin{equation}
\gamma^0 \bigl(\frac{\partial}{\partial t} +\frac{\dot \tau}{2 \tau}\bigr)
\psi - \frac{i \lambda (C -\lambda C_2)}{\tau} \gamma^0 \gamma^5 \psi =0.
\end{equation}
Defining $W(t) = \sqrt{\tau}\,\psi(t)$ one writes the foregoing equations
as
\begin{eqnarray}
\dot W_1 - \lambda \dot{\varphi} W_3 &=&0, \quad 
\dot W_2 - \lambda \dot{\varphi} W_4 = 0, \nonumber \\ 
\dot W_3 - \lambda \dot{\varphi} W_1 &=&0, \quad 
\dot W_1 - \lambda \dot{\varphi} W_3 = 0. 
\end{eqnarray}
Differentiating the fisrt equation of the foregoing system one gets
\begin{equation}
\ddot{W}_1 + \frac{\dot \tau}{\tau} \dot W_1 - [\lambda (C - \lambda C_2)]^2
\frac{1}{\tau^2}\,W_1 = 0,
\end{equation}
where the third equation of the system as well as $\dot \varphi$ has been
taken into account. The first integral of this equation reads
\begin{equation}
\tau W_1 = \lambda (C - \lambda C_2) W_1,
\end{equation}
with the constant of integration be taken trivial. Procceding analogically
one writes the solution of the system as
\begin{equation}
W_{1,3}=D_{+}\mbox{exp}\,[\lambda (C - \lambda C_2) \int \tau^{-1} dt,\quad
W_{2,4}=D_{-}\mbox{exp}\,[\lambda (C - \lambda C_2) \int \tau^{-1} dt.
\end{equation}
The components of energy-momentum tensor in this case read
$$ T_{0}^{0} = \frac{m C_0}{\tau} + \frac{{\cal C}}{\tau^2} + 
\frac{\varepsilon_0}{\tau^{1 + \xi}}, \quad 
T_{1}^{1} = T_{2}^{2} = T_{3}^{3} = -\frac{(C - \lambda C_2)^2}{2 \tau^2}
- \frac{\varepsilon_0 \xi}{\tau^{1 +\xi}},$$
where ${\cal C} := (C^2 - \lambda^2 C_{2}^{2})/2$ and
$0<\xi<1.$ Putting this into (2.25) one gets
\begin{equation}
\int\,\frac{d\tau}{\sqrt{3\kappa (m C_0 \tau + \varepsilon_0 \tau^{1-\xi} 
+{\cal C})+ {\cal X}/3}} = t.
\end{equation}
As one sees
$$ \tau|_{t \to 0} \approx \sqrt{{\cal X}/3 - 3\kappa {\cal C}}\, t \to 0,$$
and
$$ \tau|_{t \to \infty} \approx \sqrt{3 \kappa m C_0}\, t^2,$$
which means the solution obtained is initially singular and the 
isotropization process of the initially anisotropic Universe takes place
as $t \to \infty.$
\vskip 5mm

\vskip 5mm
\section{Conclusions}
                                       
\noindent
Exact solutions to the self-consistent system of spinor and scalar 
field equations have been obtained for the B-I space-time filled with
perfect fluid. It is shown that the solutions obtained are initially
singular and the space-time is basically asymptotically isotropic 
independent to the choice of interacting term in the Lagrangian, though
there are some special cases that occur initially regular (with breaking
energy-dominent condition~\cite{Alv}) solutions and leave the space-time
asymptotically anisotropic.

\end{document}